\documentclass[a4paper]{jpconf}
\usepackage{graphicx}
\begin{document}

\def\MJ{{\sc Majorana}}             
\def\DEM{{\sc Demonstrator}}             
\def\onbb{$0\nu\beta\beta$}
\def\tnbb{$2\nu\beta\beta$}
\def\onbbx{$0\nu\beta\beta\chi$}

\title{Progress Toward A \tnbb~Measurement For The \MJ~\DEM}




\newcommand{\unc}{$^{1}$}
\newcommand{\tunl}{$^{2}$}
\newcommand{\lbnl}{$^{3}$}
\newcommand{\uw}{$^{4}$}
\newcommand{\pnnl}{$^{5}$}
\newcommand{\usc}{$^{6}$}
\newcommand{\ornl}{$^{7}$}
\newcommand{\itep}{$^{8}$}
\newcommand{\usd}{$^{9}$}
\newcommand{\mpi}{$^{10}$}
\newcommand{\jinr}{$^{11}$}
\newcommand{\duke}{$^{12}$}
\newcommand{\sdsmt}{$^{13}$}
\newcommand{\lanl}{$^{14}$}
\newcommand{\ut}{$^{15}$}
\newcommand{\ou}{$^{16}$}
\newcommand{\princeton}{$^{17}$}
\newcommand{\ncsu}{$^{18}$}
\newcommand{\blhill}{$^{19}$}
\newcommand{\ttu}{$^{20}$}
\newcommand{\queens}{$^{21}$}
\newcommand{\tum}{$^{22}$}
\newcommand{\cs}{$^{,}$}
\newcommand{\clara}{$^{a}$}



\author{T~Gilliss\unc\cs\tunl, N~Abgrall\lbnl, S~I~Alvis\uw, I~J~Arnquist\pnnl, F~T~Avignone~III\usc\cs\ornl, A~S~Barabash\itep, C~J~Barton\usd, F~E~Bertrand\ornl, T~Bode\mpi, A~W~Bradley\lbnl, V~Brudanin\jinr, M~Busch\duke\cs\tunl, M~Buuck\uw, T~S~Caldwell\unc\cs\tunl, Y-D~Chan\lbnl, C~D~Christofferson\sdsmt, P~-H~Chu\lanl, C~Cuesta\uw\cs\clara, J~A~Detwiler\uw, C~Dunagan\sdsmt, Yu~Efremenko\ut\cs\ornl, H~Ejiri\ou, S~R~Elliott\lanl, G~K~Giovanetti\princeton, M~P~Green\ncsu\cs\tunl\cs\ornl, J~Gruszko\uw, I~S~Guinn\uw, V~E~Guiseppe\usc, C~R~Haufe\unc\cs\tunl, L~Hehn\lbnl, R~Henning\unc\cs\tunl, E~W~Hoppe\pnnl, M~A~Howe\unc\cs\tunl, K~J~Keeter\blhill, M~F~Kidd\ttu, S~I~Konovalov\itep, R~T~Kouzes\pnnl, A~M~Lopez\ut, R~D~Martin\queens, R~Massarczyk\lanl, S~J~Meijer\unc\cs\tunl, S~Mertens\mpi\cs\tum, J~Myslik\lbnl, C~O'Shaughnessy\unc\cs\tunl, G~Othman\unc\cs\tunl, W~Pettus\uw, A~W~P~Poon\lbnl, D~C~Radford\ornl, J~Rager\unc\cs\tunl, A~L~Reine\unc\cs\tunl, K~Rielage\lanl, R~G~H~Robertson\uw, N~W~Ruof\uw, B~Shanks\unc\cs\tunl, M~Shirchenko\jinr, A~M~Suriano\sdsmt, D~Tedeschi\usc, J~E~Trimble\unc\cs\tunl, R~L~Varner\ornl, S~Vasilyev\jinr, K~Vetter\lbnl, K~Vorren\unc\cs\tunl, B~R~White\lanl, J~F~Wilkerson\unc\cs\tunl\cs\ornl, C~Wiseman\usc, W~Xu\usd, E~Yakushev\jinr, C~-H~Yu\ornl, V~Yumatov\itep, I~Zhitnikov\jinr, and B~X~Zhu\lanl}                                                          
\address{\unc Department of Physics and Astronomy, University of North Carolina, Chapel Hill, NC, USA}
\address{\tunl Triangle Universities Nuclear Laboratory, Durham, NC, USA}
\address{\lbnl Nuclear Science Division, Lawrence Berkeley National Laboratory, Berkeley, CA, USA}
\address{\uw Center for Experimental Nuclear Physics and Astrophysics, and Department of Physics, University of Washington, Seattle, WA, USA}
\address{\pnnl Pacific Northwest National Laboratory, Richland, WA, USA}
\address{\usc Department of Physics and Astronomy, University of South Carolina, Columbia, SC, USA}
\address{\ornl Oak Ridge National Laboratory, Oak Ridge, TN, USA}
\address{\itep National Research Center ``Kurchatov Institute'' Institute for Theoretical and Experimental Physics, Moscow, Russia}
\address{\usd Department of Physics, University of South Dakota, Vermillion, SD, USA} 
\address{\mpi Max-Planck-Institut f\"{u}r Physik, M\"{u}nchen, Germany}
\address{\jinr Joint Institute for Nuclear Research, Dubna, Russia}
\address{\duke Department of Physics, Duke University, Durham, NC, USA}
\address{\sdsmt South Dakota School of Mines and Technology, Rapid City, SD, USA}
\address{\lanl Los Alamos National Laboratory, Los Alamos, NM, USA}
\address{\ut Department of Physics and Astronomy, University of Tennessee, Knoxville, TN, USA}
\address{\ou Research Center for Nuclear Physics, Osaka University, Ibaraki, Osaka, Japan}
\address{\princeton Department of Physics, Princeton University, Princeton, NJ, USA}
\address{\ncsu Department of Physics, North Carolina State University, Raleigh, NC, USA}
\address{\blhill Department of Physics, Black Hills State University, Spearfish, SD, USA}
\address{\ttu Tennessee Tech University, Cookeville, TN, USA}
\address{\queens Department of Physics, Engineering Physics and Astronomy, Queen's University, Kingston, ON, Canada} 
\address{\tum Physik Department, Technische Universit\"{a}t, M\"{u}nchen, Germany}
\vspace{0.2cm}
\address{\clara Present Address: Centro de Investigaciones Energ\'{e}ticas, Medioambientales y Tecnol\'{o}gicas, CIEMAT, 28040, Madrid, Spain}
\vspace{0.2cm}
\ead{gilliss@unc.edu}

\begin{abstract}
The \MJ~\DEM~is a $^{76}$Ge-based neutrinoless double-beta decay (\onbb) experiment. Staged at the 4850 ft level of the Sanford Underground Research Facility, the \DEM~operates an array of high-purity p-type point contact Ge detectors deployed within a graded passive shield and an active muon veto system. The present work concerns the two-neutrino double-beta decay mode (\tnbb) of $^{76}$Ge. For Ge detectors, having superior energy resolution (0.1\%), this mode poses negligible background to the \onbb~mode, even for a ton-scale experiment. However, the measurement of the \tnbb~mode allows for careful systematics checks of active detector mass, enrichment fraction, and pulse shape discrimination cuts related to both the \onbb~and \tnbb~decay modes. A precision measurement of the \tnbb~shape also allows searches for spectral distortions, possibly indicative of new physics, including \onbbx. Work is underway to construct a full experimental background model enabling a Bayesian fit to the measured energy spectrum and extraction of a precise \tnbb~spectrum and half-life.
\end{abstract}

\section{Introduction}

The \MJ~\DEM~has deployed two arrays of high-purity Ge (HPGe) crystals, 29.7 kg of which are enriched to 88\% in $^{76}$Ge. Including crystals of natural Ge abundance, a total detector mass of 44.1 kg is housed within two cryostats assembled from radiopure components, surrounded by further shielding, and operated underground \cite{MJD2014}. The \DEM~began collecting data during commissioning in June 2015, with a partial shield installation, and construction of the experiment was finished in January 2017. Background rates approaching the lowest set by \onbb~experiments have been achieved \cite{GERDATAUP17,MJDRecentResults2017}.

Central to the experimental design are HPGe crystals of p-type point contact (PPC) design. This detector technology affords excellent energy resolution, requiring only 2.96 eV per ionized electron-hole pair when operated at liquid nitrogen temperatures. Additionally, the point contact design is of low capacitance which reduces a crystal's contribution to equivalent noise charge and lowers energy thresholds in analysis. Access to the low-energy spectrum has allowed measurement of cosmogenic activation rates in Ge along with searches for candidate dark matter interactions and other exotic processes \cite{MJDDM2017}. A third advantage of the chosen detector technology arises from the shape of the electric potential within the interesting point contact geometry. The potential is peaked around the small point contact, leading to drift times of ionized charges that depend highly on the initial interaction point of radiation within the crystal bulk \cite{PPC1989,PPC2007,MJDPPC2015}. The varied pulse shapes facilitate discrimination between single-site (\onbb~signal-like) and multi-site (background-like) energy depositions within a detector based on current amplitude versus energy (AvsE)~\cite{MJDCuestaNeutrino2016}. Analysis of pulse shapes also aids in rejection of degraded alpha events incident on the passivated surface of detectors \cite{MJDGruszkoNeutrino2016,GruszkoThesis2017}. Such $\alpha$ events yield ionized charges that collect on the passivation layer and slowly drift toward the point contact in a process of delayed charge recovery (DCR).

In initial data from the \DEM, shown in Figure 1 and described in \cite{MJDRecentResults2017}, the \tnbb~spectrum is apparent both before and after removal of $\alpha$ and multi-site events through pulse shape analysis. Unlike \onbb~in which nearly all transition energy is carried by two outgoing electrons, the alternate \tnbb~decay mode includes the emission of neutrinos and is characterized by a continuous spectrum of summed electron energies ranging up to the 2039 keV $Q$-value of \onbb~($Q_{\beta\beta}$). The \tnbb~decay mode has been observed in multiple isotopes with half-lives ranging from $10^{18}$ to $10^{24}$ yr, and in $^{76}$Ge at a half-life of $1.926\times10^{21}$ yr \cite{GERDA2015}.

The \tnbb~spectrum has its endpoint at $Q_{\beta\beta}$, abutting the \onbb~region of interest (ROI) and potentially contributing events with measured energy at or near 2039 keV. Additionally, \tnbb~events share the same single-site topology as \onbb~events owing to small spatial separation of the emitted electrons within the enriched Ge. As such, \tnbb~events pose a background to \onbb~that can only be reduced through narrowed energy resolution \cite{AnRev2002}. The \DEM~has put forward the best resolution of any \onbb~experiment and reduced the potential overlap of \tnbb~and \onbb~signals to negligible levels.

\begin{figure}[h]
  \centerline{\includegraphics[width=250pt]{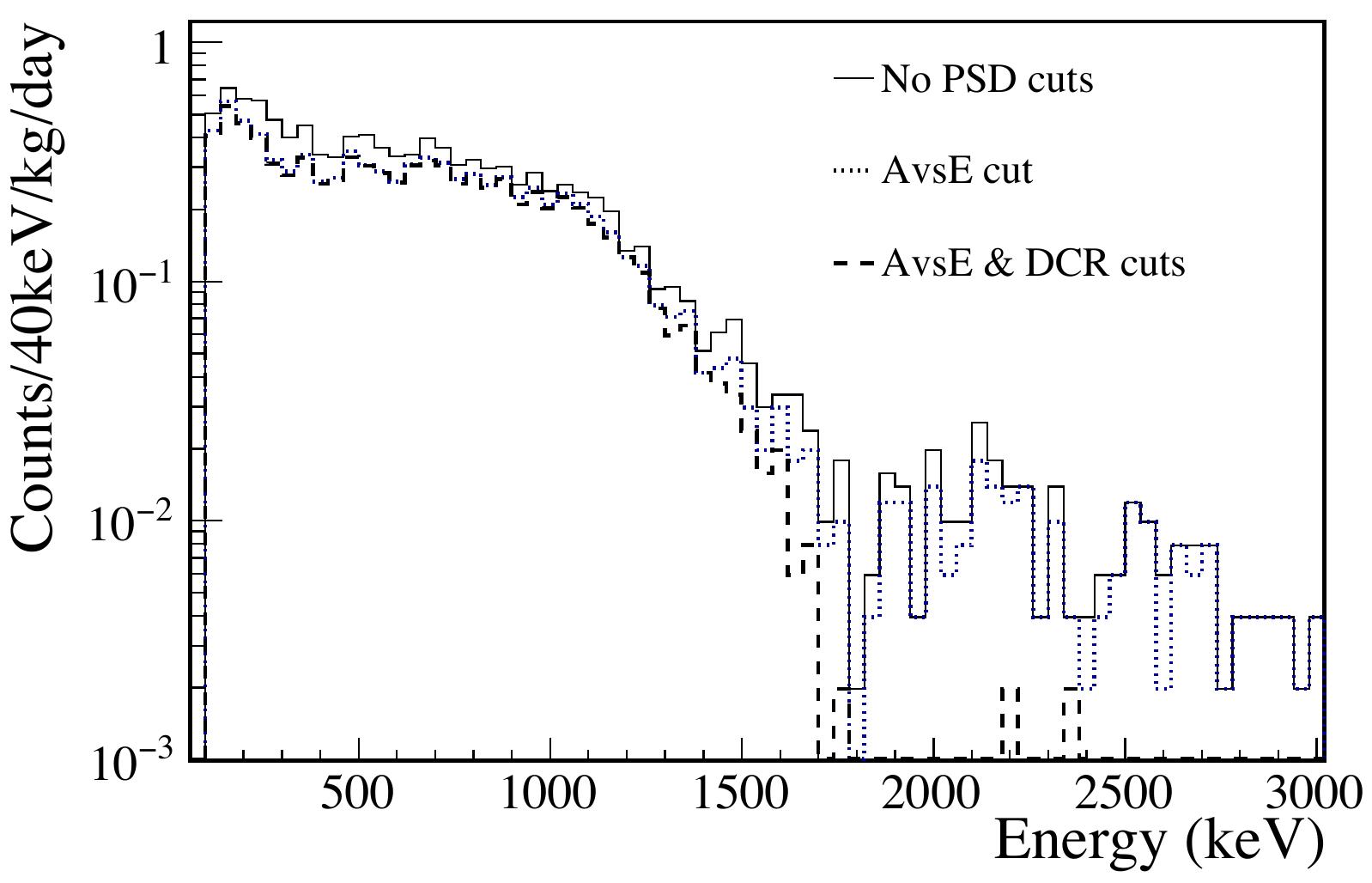}}
  \caption{Data from a 1.39 kg yr exposure of enriched detectors, shown with 40 keV binning. Multi-site and surface alpha events are removed by AvsE and DCR pulse shape cuts described in the text. With all analysis cuts in place, the enriched detector spectrum below 2039~keV is dominated by \tnbb.}
\end{figure}

With the recent achievement of near background free running of the \DEM, with respect to the \onbb~ROI, a detailed study of the apparent \tnbb~spectrum is in order. Work is underway to construct a full experimental
background model enabling a Bayesian fit to the measured energy spectrum and extraction of a precise \tnbb~half-life measurement.

\section{Backgrounds and model}

Aside from \tnbb, it is necessary to understand remaining contributions to the experimental energy spectrum. These remaining contributions are small in comparison to \tnbb~as a result of a thorough assay campaign that informed materials selection and parts processing \cite{MJDAssay2016}. As an example of the achieved radiopurity, assay of underground electroformed copper (UGEFCu) reports a combined U and Th activity of 0.23~$\mu$Bq~kg$^{-1}$, hundreds of times below the \tnbb~activity observed in enriched detectors.

Radioimpurities, like U and Th, present in the experiment's construction materials are expected as the largest contribution to \onbb~backgrounds. In response to this, the aforementioned AvsE and DCR cuts are trained to accept 90 and 99\% of good single-site events from the $^{208}$Tl 2614~keV double escape peak and Compton shoulder, respectively; the cuts then reject multi-site and surface $\alpha$ events with subsequent efficiencies. Cosmic ray muons and their secondary products also pose appreciable backgrounds to the \onbb~signal, though an active muon veto system rejects such events from the data. Additional radioimpurities accrued through cosmogenic activation are observed in experimental materials that were exposed at Earth's surface before shipment underground; time in shielded environments mitigates this class of backgrounds. Final small contributions to the background budget arise from the natural long-lived decay chains present in the rock and air external to the \DEM, and faint neutrino-electron and neutrino-nucleus scattering.

As Figure 2 summarizes, the assay and simulations campaigns have estimated the contributions of each of these background sources to the ROI. Assay techniques measure or set upper limits on the specific activity of each material, and those results are scaled based on efficiencies determined in Monte Carlo simulations that account for the detailed experimental geometry. Combining all components gives an expected background rate below $3.5$~counts~ROI$^{-1}$~t$^{-1}$~yr$^{-1}$ in a 4 keV ROI surrounding $Q_{\beta\beta}$.

%
%
\begin{figure}[h]
  \centerline{\includegraphics[width=450pt]{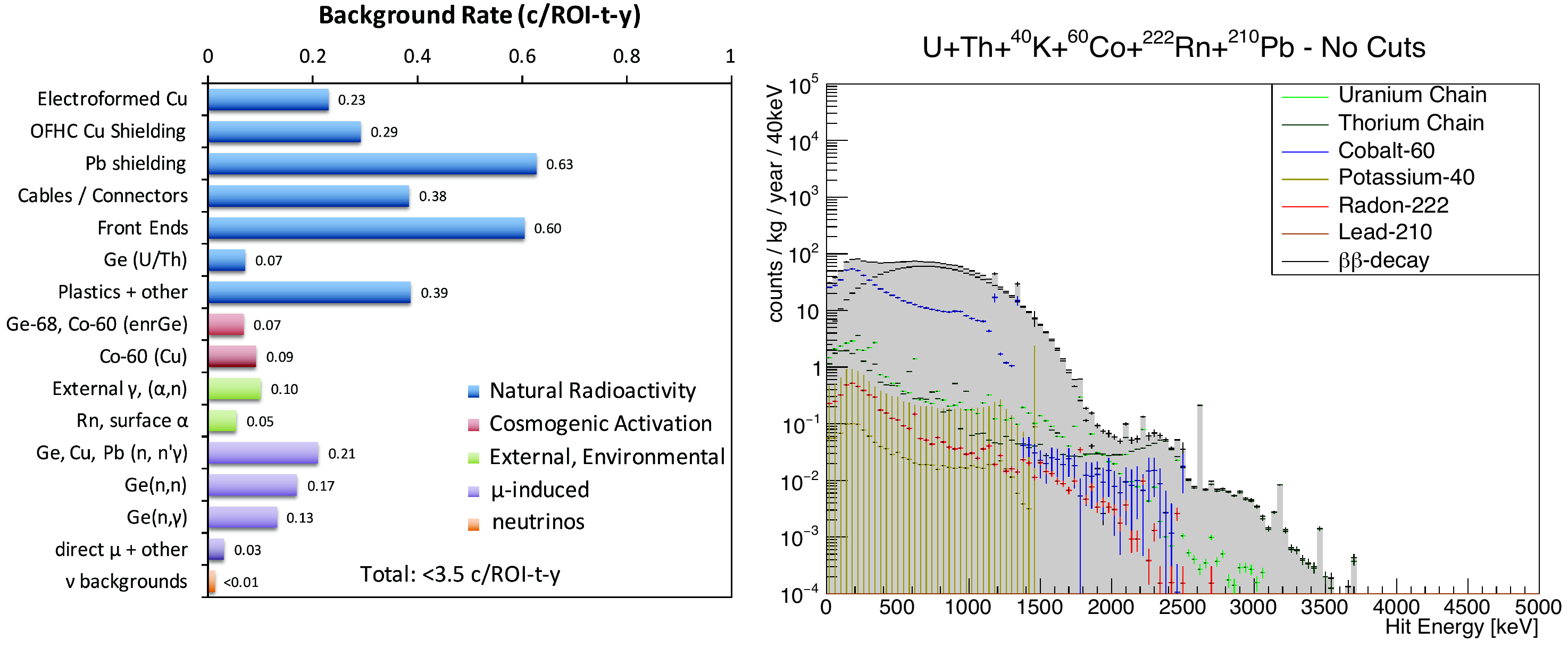}}
  \caption{(Left) Using efficiencies determined through Monte Carlo simulation, assay results are scaled to predict background contributions within a 4 keV ROI surrounding $Q_{\beta\beta}$. In the case that assay techniques report an upper limit for activity, that upper limit value is used. (Right) A model of the \DEM~background spectrum obtained from detailed Monte Carlo simulations. Key decay chains are simulated for each component of the experimental geometry and results are scaled based on materials assay. The scaled simulation results are summed in the shaded histogram.} 
\end{figure}

The combination of assay and simulation results can also provide a model for the observed energy spectrum. Geant4.10-based Monte Carlo simulations, developed in the MaGe~\cite{MAGE2011} framework and employing the full as-built geometry of the \DEM, are generated for all hardware components grouped by material. In turn, the spectral contribution of each grouping is simulated for multiple decay chains present within those materials. For instance, the background contribution of UGEFCu parts due to their Th content is simulated, and subsequent jobs cover their U, K, and Co contents.

From assay, simulation, and analysis of real data, estimates of the most significant background components can be made. By summing the contributions of the most significant components, the spectrum from Monte Carlo simulations is in qualitative agreement with the experimental spectrum. Figure 2 provides an example.
%

\section{Overview of \tnbb~measurement}

With a detailed background model and growing exposure, the \tnbb~rate can be extracted from a fit of the model to the measured spectrum. A large set of parameters describes the experimental energy spectrum observed by each of the \DEM~detectors, including intensities of background components, detector mass, active volume, and detection efficiency. Given the number of free parameters in such a multi-detector model, a best fit to the values for each detector will be found through a Markov chain Monte Carlo procedure. Detector-specific parameters are allowed to float independently while global parameters, like \tnbb~rate, are fixed for the array.

Some of the parameter space, regarding source location and intensity of background components, can be constrained through spectral analysis. For instance, background components making only faint contributions, and which cannot be unambiguously identified, could be omitted. In a Bayesian implementation of the fit, prior distributions can be set to further constrain background component intensities, detector characteristics, and the \tnbb~half-life.

Given the low-background design of the \DEM, the low-statistics data poses a challenge to identifying sub-dominant background components, especially those lacking characteristic peaks. Multiply scattered events and degraded alpha events, mentioned above, can take on a continuum of energies and fall into this category. As a countermeasure, simulations and analysis work can compare the results of individual detector fits and higher-statistics fits, involving all operating detectors, to help constrain backgrounds sources based on simulations. AvsE and DCR cuts can also be applied to reduce the presence of continuum backgrounds so long as the uncertainties associated with these cuts are accounted for in any results.

With a functional background model in place, additional searches can take place for \onbb~decay modes involving emission of Majorons (\onbbx). These decay modes alter the outgoing phase space of \onbb~and lead to variations in the shape of the electrons' summed energy spectrum. Searches for Majoron-emitting double-beta decays will proceed by including a \onbbx~component in the spectral fit, with its amplitude floating alongside those of the components of the background model. Limits on the prevalence of other new physics can also be set based on distortions of spectral shape.

\section{Outlook}

To date, the \MJ~\DEM~has accumulated an appreciable exposure of HPGe detectors under low-background conditions, and a simulation campaign informed by materials assay has studied the transport of radiation through a detailed model of the experimental setup. Work is ongoing to constrain the location and importance of various background contributions, with insights guiding a full fit to the experimental energy spectrum. Additional work discussed above will determine the \tnbb~half-life and search for Majoron-emitting double-beta decays, studies which are complimentary to a \onbb~analysis.

\section{Acknowledgements}
This material is based upon work supported by the U.S. Department of Energy, Office of Science, Office of Nuclear
Physics, the Particle Astrophysics and Nuclear Physics Programs of the National Science Foundation, and the Sanford
Underground Research Facility.



\end{document}